\documentclass[preprint,12pt]{elsarticle}
\usepackage{amsmath,amssymb}
\usepackage{graphicx}
\usepackage{booktabs}
\usepackage[margin=2.2cm]{geometry}
\usepackage[colorlinks=true,linkcolor=blue,citecolor=blue,urlcolor=blue]{hyperref}

\makeatletter
\def\ps@pprintTitle{\let\@oddhead\@empty\let\@evenhead\@empty\def\@oddfoot{\centerline{\thepage}}\let\@evenfoot\@oddfoot}
\makeatother

\begin{document}

\begin{frontmatter}

\title{Holographic dark energy through Kruglov entropy}

\author[gla]{Umesh Kumar Sharma}
\ead{sharma.umesh@gla.ac.in}
\author[utas]{P. Suresh Kumar\corref{cor1}}
\ead{sureshharinirahav@gmail.com}
\cortext[cor1]{Corresponding author}
\address[gla]{Department of Mathematics, Institute of Applied Sciences and Humanities, GLA University, Mathura, Uttar Pradesh 281406, India}
\address[utas]{Department of Supportive Requirements, University of Technology and Applied Sciences-Salalah, Salalah, Sultanate of Oman}

\begin{abstract}
We study a holographic dark energy model built on the new entropy proposed by Kruglov, $S_K=S_{BH}/(1+\gamma S_{BH}^2)$, with the Granda--Oliveros infrared cutoff $L^{-2}=\alpha H^2+\beta\dot H$ in a flat universe. The Kruglov entropy is smaller than the Bekenstein--Hawking entropy and has a maximum value. We obtain the energy density of the model, the differential equation for the dark energy density parameter, and the expressions for the equation of state parameter, the deceleration parameter and the squared sound speed. At early times the model behaves like the usual Granda--Oliveros holographic dark energy, and the dark energy follows the background with a constant fraction. At late times the Kruglov correction gives a stable de Sitter future when $\alpha>1$, with the final Hubble rate fixed by the entropy parameter. The model gives a matter era followed by a dark energy era, with a transition from deceleration to acceleration near $z\approx0.66$. For larger values of the entropy parameter the equation of state can cross the phantom divide; for the parameters allowed by the data this happens only in the future, if at all. We constrain the model with 31 cosmic chronometer points, Pantheon+ supernovae and DESI DR2 baryon acoustic oscillation data, together with a big bang nucleosynthesis condition, using a Markov chain Monte Carlo analysis. We find $H_0=68.0\pm1.6$ km\,s$^{-1}$\,Mpc$^{-1}$, $\alpha=1.19^{+0.10}_{-0.08}$, $\beta=0.61^{+0.06}_{-0.05}$ and the 95\% upper limit $\tilde b=b/H_0^4<0.24$ on the dimensionless Kruglov parameter (for a matter-era dark energy fraction below 0.3), which gives $\gamma<4.9\times10^{-246}$. The present equation of state is $w_{D0}=-0.83\pm0.03$. The model fits the data slightly better than $\Lambda$CDM ($\Delta\chi^2=-4.5$) and is equally good by the Akaike criterion, while the Bayesian criterion and the Bayesian evidence prefer $\Lambda$CDM.
\end{abstract}


\end{frontmatter}

\section{Introduction}
\label{sec:intro}

Observations of type Ia supernovae (SNIa) first showed that the expansion of the universe is accelerating at late times \cite{Riess1998,Perlmutter1999}. Later, the cosmic microwave background (CMB) and baryon acoustic oscillation (BAO) data confirmed this result \cite{Planck2018,Eisenstein2005}. The simplest way to explain the acceleration is the cosmological constant $\Lambda$. The resulting $\Lambda$CDM model fits most data well. But it has the fine tuning and the coincidence problems \cite{Weinberg1989,Copeland2006}. Also, the recent BAO data of the Dark Energy Spectroscopic Instrument (DESI DR2), when combined with supernova and CMB data, show a mild preference for a dark energy (DE) that changes with time \cite{DESI2025}. For these reasons, many dynamical DE models have been studied.

Holographic dark energy (HDE) is one such model. It comes from the holographic principle \cite{tHooft1993,Susskind1995}. Cohen et al. \cite{Cohen1999} argued that the total energy in a region of size $L$ should not be larger than the mass of a black hole of the same size. This gives $\rho_D L^4 \lesssim S$, where $S$ is the entropy of the horizon. With the Bekenstein--Hawking (BH) entropy $S_{BH}=A/(4G)$ \cite{Bekenstein1973,Hawking1975}, one gets the standard HDE density $\rho_D=3c^2M_p^2L^{-2}$ \cite{Li2004,Hsu2004,Wang2017}. Here $M_p$ is the reduced Planck mass, $c$ is a constant and $L$ is the infrared (IR) cutoff. The HDE model therefore depends on two choices, namely the horizon entropy and the IR cutoff.

The BH entropy is not the only possible choice. Gravity is a long range force, and for such systems non-additive entropies may be more suitable. This idea gave the Tsallis \cite{Tsallis2013,Tavayef2018}, Barrow \cite{Barrow2020,Saridakis2020}, R\'enyi \cite{Moradpour2018}, Sharma--Mittal \cite{Jahromi2018,Naeem2024} and Kaniadakis \cite{Kaniadakis2002,Drepanou2022,Sheykhi2024} forms of HDE and of entropic cosmology. More general entropies with several parameters have also been proposed \cite{Nojiri2022a,Nojiri2022b,Luciano2026a}. In our earlier works we studied HDE with Kaniadakis entropy in a non-flat universe \cite{Kumar2023} and the new Tsallis HDE \cite{Pandey2022,Sharma2026}. Recently, the DESI DR2 data were used to test the Barrow and Tsallis HDE models \cite{Luciano2026b}.

In 2025, Kruglov proposed a new one-parameter entropy of the apparent horizon \cite{Kruglov2025},
\begin{equation}
S_K=\frac{S_{BH}}{1+\gamma S_{BH}^2},
\label{eq:SK}
\end{equation}
where $\gamma$ is a positive parameter. For $\gamma\to 0$, Eq.~(\ref{eq:SK}) gives back the BH entropy. For small $\gamma$ we have \cite{Kruglov2025}
\begin{equation}
S_K=S_{BH}-\gamma S_{BH}^3+\gamma^2 S_{BH}^5+\mathcal{O}(\gamma^3).
\label{eq:SKexp}
\end{equation}
So the correction decreases the BH entropy. This is different from the Kaniadakis entropy, $S=\sinh(KS_{BH})/K=S_{BH}+K^2S_{BH}^3/6+\cdots$, whose correction is positive \cite{Drepanou2022}. Kruglov noted that a negative correction is similar to what is expected from quantum gravity \cite{Kruglov2025}. Another property of Eq.~(\ref{eq:SK}) is that, as a function of $S_{BH}$, $S_K$ has a maximum value $S_K^{\rm max}=1/(2\sqrt{\gamma})$ at $S_{BH}=1/\sqrt{\gamma}$, and $S_K\to0$ when $S_{BH}\to\infty$. The Tsallis, Barrow, R\'enyi and Kaniadakis entropies have no such maximum. (In Ref.~\cite{Kruglov2025} $S_K$ is called monotonically increasing; this is true on the branch $\gamma S_{BH}^2<1$, which is the branch relevant for us, see Section~\ref{sec:obs}.) Using Eq.~(\ref{eq:SK}) with apparent horizon thermodynamics, Kruglov obtained modified Friedmann equations with a dynamical cosmological constant \cite{Kruglov2025}. The big bang nucleosynthesis (BBN) bound on this entropy was obtained in Ref.~\cite{Phukan2026}. Another Kruglov-type entropy, $S_{BH}/(1+\gamma S_{BH})$, was recently tested with DESI DR2 data in the apparent horizon approach \cite{Muhammad2026}. To our knowledge, the entropy (\ref{eq:SK}) has not yet been used in the holographic relation $\rho_DL^4\propto S$ with the GO cutoff, nor tested with DESI DR2 data.

For the IR cutoff, the Hubble horizon does not give acceleration in the simple HDE \cite{Hsu2004}, and the future event horizon brings a causality problem \cite{Li2004,Wang2017}. Granda and Oliveros (GO) proposed a local cutoff $L^{-2}=\alpha H^2+\beta\dot H$ \cite{Granda2008,Granda2009}. It depends only on the local quantities $H$ and $\dot H$, so it has no causality problem. It also contains the Ricci scale as a special case \cite{Gao2009}. The GO cutoff has been used with Barrow entropy \cite{Oliveros2022,Motaghi2024}, with Barrow and Tsallis entropies and DESI data \cite{Luciano2026c}, with multi-parameter entropies \cite{Chen2026} and with the new Tsallis entropy \cite{Sharma2026}.

In this paper, we combine the Kruglov entropy (\ref{eq:SK}) with the GO cutoff in a flat universe. We call this model Kruglov holographic dark energy (KrHDE). We obtain the differential equation for the DE density parameter, and the expressions for the equation of state (EoS) parameter, the deceleration parameter and the squared sound speed. We study the early and late time behaviour of the model. Then we constrain the model parameters with cosmic chronometer (CC), Pantheon+ supernova and DESI DR2 BAO data, and compare the result with $\Lambda$CDM and with the GO-HDE model built on the BH entropy.

The paper is arranged as follows. In Section~\ref{sec:model} we construct the model. In Section~\ref{sec:evol} we discuss the cosmological evolution. Section~\ref{sec:obs} gives the observational constraints. Section~\ref{sec:summary} gives the summary of results.

\section{Kruglov holographic dark energy with GO cutoff}
\label{sec:model}

\subsection{Energy density}

For a horizon of radius $L$, the BH entropy is $S_{BH}=\pi L^2/G$. In the standard HDE, $\rho_D\propto S_{BH}L^{-4}$, which gives $\rho_D=3c^2M_p^2L^{-2}$. Following the method used for the Tsallis, R\'enyi, Barrow and Kaniadakis HDE models \cite{Tavayef2018,Moradpour2018,Saridakis2020,Drepanou2022}, we use the Kruglov entropy (\ref{eq:SK}) in the holographic relation $\rho_D L^4\propto S_K$. Keeping the same normalisation as the standard HDE, this means $\rho_D=3c^2M_p^2L^{-2}\,(S_K/S_{BH})$. From Eq.~(\ref{eq:SK}), $S_K/S_{BH}=1/(1+\gamma S_{BH}^2)$, and with $S_{BH}^2=\pi^2L^4/G^2$ we get
\begin{equation}
\rho_D=\frac{3c^2M_p^2}{L^2}\,\frac{1}{1+bL^4},\qquad b=\frac{\pi^2\gamma}{G^2},
\label{eq:rhoL}
\end{equation}
where $b$ has the dimension of $H^4$. We use the same symbol $b$ as in Ref.~\cite{Kruglov2025}. For $b=0$ we get the standard HDE. Differentiating Eq.~(\ref{eq:SK}) gives $dS_K/dL=(dS_{BH}/dL)\,(1-bL^4)/(1+bL^4)^2$, so the entropy increases with $L$ only when $bL^4<1$, i.e. $\gamma S_{BH}^2<1$.

For the IR cutoff we take the GO form \cite{Granda2008,Granda2009},
\begin{equation}
L^{-2}=\alpha H^2+\beta\dot H,
\label{eq:GO}
\end{equation}
where $\alpha$ and $\beta$ are constants. The constant $c^2$ is absorbed into $\alpha$ and $\beta$ (and $b$ is rescaled accordingly), so we set $c=1$. Then Eq.~(\ref{eq:rhoL}) becomes
\begin{equation}
\rho_D=3M_p^2\,\frac{(\alpha H^2+\beta\dot H)^3}{(\alpha H^2+\beta\dot H)^2+b}.
\label{eq:rhoGO}
\end{equation}
For $b=0$, Eq.~(\ref{eq:rhoGO}) reduces to the GO-HDE density $\rho_D=3M_p^2(\alpha H^2+\beta\dot H)$ \cite{Granda2008}. For $\alpha=2\beta$, the cutoff (\ref{eq:GO}) is proportional to the Ricci scalar $R=6(\dot H+2H^2)$, which gives the Ricci dark energy \cite{Gao2009}.

\subsection{Friedmann equations and the density parameter}

We consider a flat Friedmann--Lema\^itre--Robertson--Walker (FLRW) universe,
\begin{equation}
ds^2=-dt^2+a^2(t)\left(dr^2+r^2d\Omega^2\right),
\end{equation}
filled with pressureless matter (dark matter and baryons) and the KrHDE. At late times the radiation is very small, so we neglect it in this section. The Friedmann equations are \cite{Copeland2006,Wang2017}
\begin{eqnarray}
3M_p^2H^2&=&\rho_m+\rho_D, \label{eq:F1}\\
2M_p^2\dot H+3M_p^2H^2&=&-p_D, \label{eq:F2}
\end{eqnarray}
where $H=\dot a/a$ and $p_D$ is the pressure of DE. There is no interaction between the two sectors, so
\begin{eqnarray}
\dot\rho_m+3H\rho_m&=&0, \label{eq:cm}\\
\dot\rho_D+3H(1+w_D)\rho_D&=&0, \label{eq:cd}
\end{eqnarray}
with $w_D=p_D/\rho_D$. The density parameters $\Omega_m=\rho_m/(3M_p^2H^2)$ and $\Omega_D=\rho_D/(3M_p^2H^2)$ satisfy $\Omega_m+\Omega_D=1$.

We use $x=\ln a$ as the time variable, with $a_0=1$ today, so that $x=-\ln(1+z)$. A prime denotes $d/dx$, and $E=H/H_0$. Since $\dot H=H_0^2 EE'=\frac{1}{2}H_0^2(E^2)'$, we define the dimensionless cutoff and the dimensionless entropy parameter as
\begin{equation}
u\equiv\frac{\alpha H^2+\beta\dot H}{H_0^2}=\alpha E^2+\frac{\beta}{2}(E^2)',\qquad \tilde b\equiv\frac{b}{H_0^4}.
\label{eq:u}
\end{equation}
Inserting Eq.~(\ref{eq:GO}) in Eq.~(\ref{eq:rhoGO}) and dividing numerator and denominator by $H_0^4$, we have $\rho_D=3M_p^2H_0^2\,u^3/(u^2+\tilde b)$. Dividing by $3M_p^2H^2=3M_p^2H_0^2E^2$, the DE density parameter is
\begin{equation}
\Omega_D=\frac{u^3}{E^2\left(u^2+\tilde b\right)},
\label{eq:OmD}
\end{equation}
Since $\rho_m=3M_p^2H_0^2\,\Omega_{m0}e^{-3x}$ from Eq.~(\ref{eq:cm}), the first Friedmann equation (\ref{eq:F1}) divided by $3M_p^2H_0^2$ takes the form
\begin{equation}
E^2=\Omega_{m0}e^{-3x}+\frac{u^3}{u^2+\tilde b}.
\label{eq:E2}
\end{equation}
For a given $E^2$, multiplying Eq.~(\ref{eq:E2}) by $(u^2+\tilde b)$ gives a cubic equation for $u$,
\begin{equation}
u^3-\mathcal{D}u^2-\mathcal{D}\tilde b=0,\qquad \mathcal{D}\equiv\Omega_DE^2=E^2-\Omega_{m0}e^{-3x}.
\label{eq:cubic}
\end{equation}
The function $u^3/(u^2+\tilde b)$ is increasing in $u$, so Eq.~(\ref{eq:cubic}) has only one real root. By Cardano's method (the steps are given in Appendix~\ref{app:cardano}) it is
\begin{equation}
u=\frac{\mathcal{D}}{3}+\mathcal{A}+\frac{\mathcal{D}^2}{9\mathcal{A}},\qquad
\mathcal{A}=\left[\frac{\mathcal{D}^3}{27}+\frac{\mathcal{D}\tilde b}{2}+\sqrt{\frac{\mathcal{D}^4\tilde b}{27}+\frac{\mathcal{D}^2\tilde b^2}{4}}\,\right]^{1/3}.
\label{eq:root}
\end{equation}
For $\tilde b=0$ we get $\mathcal{A}=\mathcal{D}/3$ and $u=\mathcal{D}$, as expected.

Now we derive the evolution equation for $\Omega_D$. From the definition (\ref{eq:u}) we have
\begin{equation}
(E^2)'=\frac{2}{\beta}\left(u-\alpha E^2\right).
\label{eq:E2p}
\end{equation}
From $\Omega_m=1-\Omega_D$ and Eq.~(\ref{eq:cm}), we get
\begin{equation}
E^2=\frac{\Omega_{m0}e^{-3x}}{1-\Omega_D},
\label{eq:E2Om}
\end{equation}
and taking the $x$-derivative gives $(E^2)'/E^2=\Omega_D'/(1-\Omega_D)-3$. Using this in Eq.~(\ref{eq:E2p}), we obtain
\begin{equation}
\Omega_D'=(1-\Omega_D)\left[3-\frac{2\alpha}{\beta}+\frac{2}{\beta}\,\frac{u\,(1-\Omega_D)}{\Omega_{m0}e^{-3x}}\right],
\label{eq:OmDp}
\end{equation}
where $u$ is given by Eq.~(\ref{eq:root}) with $\mathcal{D}=\Omega_{m0}e^{-3x}\,\Omega_D/(1-\Omega_D)$. Eq.~(\ref{eq:OmDp}) describes the evolution of KrHDE. It depends on $x$ explicitly through $\mathcal{D}$, so we solve it numerically with the initial condition $\Omega_D(x=0)=\Omega_{D0}=1-\Omega_{m0}$. For $\tilde b=0$ we have $u(1-\Omega_D)/(\Omega_{m0}e^{-3x})=\Omega_D$, and Eq.~(\ref{eq:OmDp}) becomes
\begin{equation}
\Omega_D'=(1-\Omega_D)\left[3-\frac{2\alpha}{\beta}+\frac{2\Omega_D}{\beta}\right],
\end{equation}
which is the equation of GO-HDE in a flat universe (see Refs.~\cite{Oliveros2022,Motaghi2024} in the limit of BH entropy). Thus the Kruglov parameter $\tilde b$ enters only through the root $u$.

\subsection{Equation of state, deceleration parameter and sound speed}

Adding Eqs.~(\ref{eq:cm}) and (\ref{eq:cd}) gives $\dot\rho_m+\dot\rho_D=-3H(\rho_m+\rho_D+p_D)$. Using Eq.~(\ref{eq:F1}), the left side is $6M_p^2H\dot H$, so $2\dot H/H^2=-3(1+p_D/(\rho_m+\rho_D))$. Since $\dot H/H^2=(E^2)'/(2E^2)$ and $p_D/(\rho_m+\rho_D)=w_D\Omega_D$, this becomes $(E^2)'/E^2=-3(1+w_D\Omega_D)$. Solving for $w_D$ and using Eq.~(\ref{eq:E2p}) for $(E^2)'$, the EoS parameter of KrHDE is
\begin{equation}
w_D=-\frac{1}{\Omega_D}\left[1+\frac{2}{3\beta}\left(\frac{u}{E^2}-\alpha\right)\right]=-\frac{\Omega_D'}{3\Omega_D(1-\Omega_D)}.
\label{eq:wD}
\end{equation}
The second form follows from $(E^2)'/E^2=\Omega_D'/(1-\Omega_D)-3$. For $\tilde b=0$, $u/E^2=\Omega_D$ and Eq.~(\ref{eq:wD}) gives $w_D=(2\alpha-3\beta-2\Omega_D)/(3\beta\Omega_D)$, which is the GO-HDE result \cite{Oliveros2022,Motaghi2024}. The deceleration parameter follows from $\dot H/H^2=(E^2)'/(2E^2)$ and Eq.~(\ref{eq:E2p}):
\begin{equation}
q=-1-\frac{\dot H}{H^2}=-1-\frac{(E^2)'}{2E^2}=\frac{1}{\beta}\left(\alpha-\beta-\frac{u}{E^2}\right)=\frac{1}{2}\left(1+3w_D\Omega_D\right).
\label{eq:q}
\end{equation}
To check the classical stability of the model, we use the squared sound speed \cite{Myung2007,Kim2008},
\begin{equation}
v_s^2=\frac{dp_D}{d\rho_D}=\frac{p_D'}{\rho_D'}=w_D+\frac{\rho_D}{\rho_D'}\,w_D'=w_D-\frac{w_D'}{3(1+w_D)},
\label{eq:vs2}
\end{equation}
where we used $p_D=w_D\rho_D$ and $\rho_D'/\rho_D=-3(1+w_D)$ from Eq.~(\ref{eq:cd}). The model is stable against small perturbations when $v_s^2>0$. To find $w_D'$ we differentiate Eq.~(\ref{eq:wD}). Writing $w_D=-\left[E^2+(E^2)'/3\right]/\mathcal{D}$, we need $\mathcal{D}'=(E^2)'+3\Omega_{m0}e^{-3x}$ and $(E^2)''=(2/\beta)\left(u'-\alpha(E^2)'\right)$. To find $u'$, we differentiate the cubic (\ref{eq:cubic}): $(3u^2-2\mathcal{D}u)u'=\mathcal{D}'(u^2+\tilde b)$. Using $\mathcal{D}=u^3/(u^2+\tilde b)$, we have $3u-2\mathcal{D}=u(u^2+3\tilde b)/(u^2+\tilde b)$, and so
\begin{equation}
u'=\mathcal{D}'\,\frac{(u^2+\tilde b)^2}{u^2(u^2+3\tilde b)}.
\end{equation}

\subsection{Early and late time behaviour}
\label{sec:limits}

At early times $H\gg H_0$ and so $u\gg1$. Then $\tilde b/u^2\propto(H_0/H)^4$ is very small, and the KrHDE behaves like the GO-HDE. If the background evolves as $E^2\propto a^{-3(1+w_b)}$, then $(E^2)'/E^2=-3(1+w_b)$. Putting this in the definition (\ref{eq:u}) gives $u/E^2=\alpha-\frac{3}{2}\beta(1+w_b)$, and since $u\gg\sqrt{\tilde b}$, Eq.~(\ref{eq:OmD}) gives $\Omega_D\simeq u/E^2$. So the DE follows the background with a constant fraction (the same tracking relations were found for the new Tsallis HDE with GO cutoff in Ref.~\cite{Sharma2026}),
\begin{equation}
\Omega_D^{(r)}=\alpha-2\beta\quad(\text{radiation era}),\qquad
\Omega_D^{(m)}=\alpha-\frac{3}{2}\beta\quad(\text{matter era}).
\label{eq:track}
\end{equation}
These fractions do not depend on $\tilde b$. A positive DE density in the matter era needs $\Omega_D^{(m)}\geq0$. In the radiation era, a non-zero $\Omega_D^{(r)}$ changes the expansion rate at BBN like extra neutrino species. Before $e^\pm$ annihilation the relativistic degrees of freedom are $g_*=43/4$, and one neutrino species contributes $7/4$ \cite{Kolb1990}. So $\rho_D/\rho_r=\Omega_D^{(r)}/(1-\Omega_D^{(r)})$ gives \cite{Sharma2026}
\begin{equation}
\Delta N_\nu=\frac{43}{7}\,\frac{\Omega_D^{(r)}}{1-\Omega_D^{(r)}}.
\label{eq:dNnu}
\end{equation}
We will use this relation in Section~\ref{sec:obs}. Note that the BBN bound on the Kruglov parameter in Ref.~\cite{Phukan2026} was obtained for the apparent horizon cosmology of Ref.~\cite{Kruglov2025}, not for the present HDE model. In our model the Kruglov term is negligible at BBN, because $\tilde b/u^2\propto(H_0/H)^4$.

At late times the Kruglov term becomes important, because $u$ decreases. When matter dilutes away, $\Omega_D\to1$ and Eq.~(\ref{eq:E2p}) becomes $(E^2)'=(2/\beta)\left(u-\alpha E^2\right)$ with $E^2=u^3/(u^2+\tilde b)$. A de Sitter solution needs $(E^2)'=0$, i.e. $u=\alpha E^2$. Putting $u=\alpha E^2$ in $E^2=u^3/(u^2+\tilde b)$ gives $\alpha^2E^4+\tilde b=\alpha^3E^4$, so
\begin{equation}
E_{dS}^4=\frac{H_{dS}^4}{H_0^4}=\frac{\tilde b}{\alpha^2(\alpha-1)},
\label{eq:dS}
\end{equation}
which exists for $\alpha>1$. To check its stability, we write the late-time equation as $(E^2)'=F(E^2)\equiv(2/\beta)\left[u(E^2)-\alpha E^2\right]$, with $du/dE^2=(u^2+\tilde b)^2/[u^2(u^2+3\tilde b)]$ (the same relation as for $u'$ above). At the de Sitter point $u^2=\alpha^2E_{dS}^4=\tilde b/(\alpha-1)$, so $u^2+\tilde b=\alpha\tilde b/(\alpha-1)$ and $u^2+3\tilde b=(3\alpha-2)\tilde b/(\alpha-1)$. This gives $du/dE^2=\alpha^2/(3\alpha-2)$ and
\begin{equation}
\frac{dF}{dE^2}=\frac{2}{\beta}\left[\frac{\alpha^2}{3\alpha-2}-\alpha\right]=-\frac{4\alpha(\alpha-1)}{\beta(3\alpha-2)},
\end{equation}
which is negative for $\alpha>1$ and $\beta>0$. A small deviation $\delta$ from $E_{dS}^2$ then obeys $\delta'=(dF/dE^2)\,\delta$ and decays. Hence the de Sitter point is a stable future attractor. GO-HDE does not have this feature. In the GO-HDE ($\tilde b=0$) with $\alpha>1$, the Hubble parameter goes to zero in the far future, and for $\alpha<1$ it grows without limit (phantom behaviour). In KrHDE with $\alpha>1$, the final Hubble rate is fixed by the entropy parameter through Eq.~(\ref{eq:dS}). At the de Sitter point $u^2=\tilde b/(\alpha-1)$. Since $L^{-2}=H_0^2u$, we have $bL^4=\tilde b/u^2$, so the entropy increases with $L$ only for $u^2>\tilde b$. Therefore the final state lies on the increasing part of $S_K$ when $1<\alpha<2$.

\section{Cosmological evolution}
\label{sec:evol}

In this section we study the behaviour of $\Omega_D$, $w_D$, $q$ and $v_s^2$. Eq.~(\ref{eq:OmDp}) has no analytic solution for $\tilde b\neq0$, so we solve it numerically. We fix $\Omega_{m0}=0.223$, $\alpha=1.07$ and $\beta=0.547$, which are the best fit values of KrHDE found in Section~\ref{sec:obs}. The KrHDE best fit has $\tilde b\simeq0$, so it is the same point as the GO-HDE best fit. These best fit values differ a little from the medians in Table~\ref{tab:post} (e.g. $\Omega_{m0}=0.212$), because the medians are averaged over the allowed range of $\tilde b$ and are not one consistent point. We take $\tilde b=0.1$, inside the 95\% range, as the reference value. Note that $\Omega_{m0}=0.223$ is lower than the usual value 0.3. This is because:
\begin{itemize}
\item with the GO cutoff, about 25\% of the DE behaves like matter at high redshift, $\Omega_D^{(m)}=\alpha-3\beta/2\approx0.25$ (Section~\ref{sec:limits});
\item the data measure the total that behaves like matter, which is the effective matter density $\Omega_{m0}/(1-\Omega_D^{(m)})\approx0.223/0.75\approx0.30$. This is the usual value (see Section~\ref{sec:results}).
\end{itemize} In every figure, panel (a) shows the effect of $\tilde b$, panel (b) the effect of $\alpha$ and panel (c) the effect of $\beta$, while the other parameters are kept fixed. The curve with $\tilde b=0$ in panel (a) is the GO-HDE. The plots cover $-1<z\le3$, so they include the future ($z<0$). For these values, $\Omega_D^{(m)}=\alpha-3\beta/2\approx0.25$ and $\Omega_D^{(r)}=\alpha-2\beta\approx-0.02$.

\begin{figure}[!htbp]
\centering
\includegraphics[width=\textwidth]{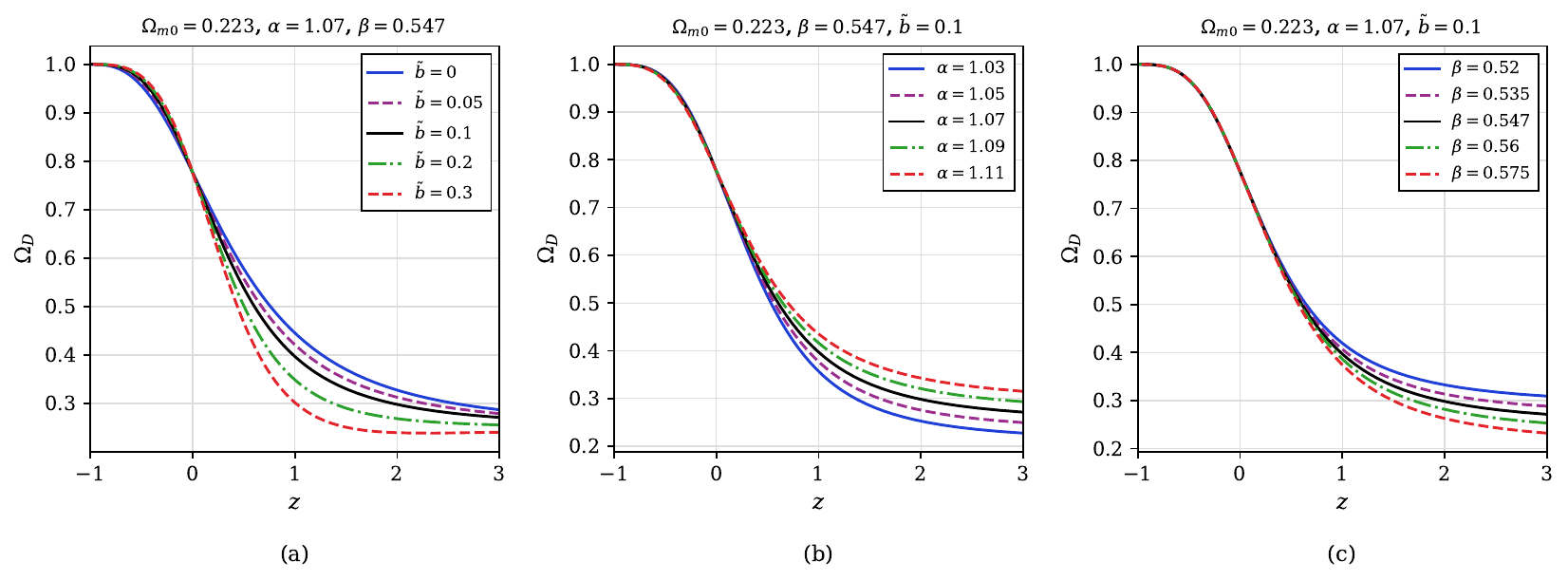}
\caption{The evolution of the KrHDE density parameter $\Omega_D$ with redshift $z$. (a) Different $\tilde b$, (b) different $\alpha$, (c) different $\beta$. The fixed values are written above each panel.}
\label{fig:Omd}
\end{figure}

Fig.~\ref{fig:Omd} shows $\Omega_D(z)$. At high redshift $\Omega_D$ changes slowly and approaches the matter-era value $\Omega_D^{(m)}$ of Eq.~(\ref{eq:track}). At low redshift $\Omega_D$ grows quickly, and in the far future ($z\to-1$) the universe becomes fully dominated by DE, $\Omega_D\to1$. So the model gives a matter dominated past and a DE dominated future. Increasing $\tilde b$ lowers $\Omega_D$ in the range $0<z<3$. The reason is that the Kruglov correction reduces the entropy, and so it reduces the holographic density at a given $L$. The value $\Omega_D(z=3)$ falls from about 0.29 for $\tilde b=0$ to about 0.24 for $\tilde b=0.3$. Panels (b) and (c) show that a larger $\alpha$ or a smaller $\beta$ raises $\Omega_D$ in the past, as expected from $\Omega_D^{(m)}=\alpha-3\beta/2$.

\begin{figure}[!htbp]
\centering
\includegraphics[width=\textwidth]{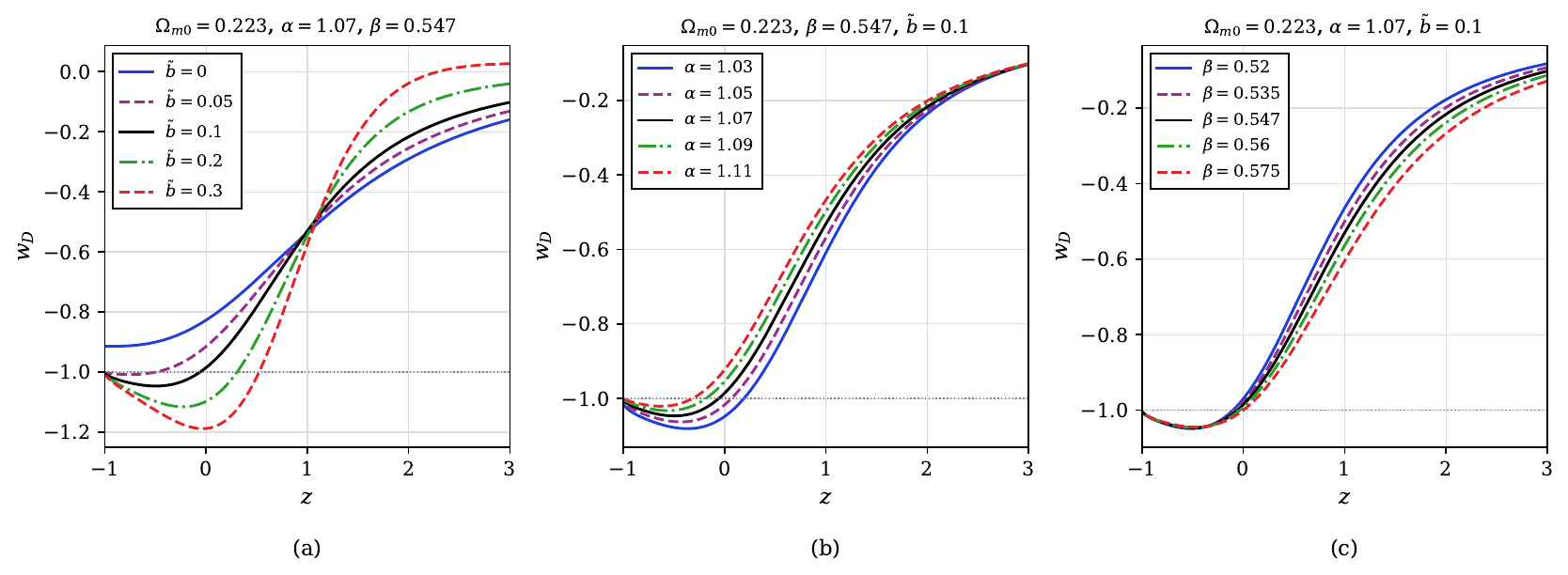}
\caption{The evolution of the KrHDE EoS parameter $w_D$ with redshift $z$. (a) Different $\tilde b$, (b) different $\alpha$, (c) different $\beta$. The dotted line is $w_D=-1$.}
\label{fig:wD}
\end{figure}

The EoS parameter is plotted in Fig.~\ref{fig:wD}. For $\tilde b=0$ (GO-HDE), $w_D$ stays above $-1$ at all times, so the DE is of quintessence type. Today $w_D\approx-0.83$, and in the far future $w_D\to-1+2(\alpha-1)/(3\beta)\approx-0.92$ \cite{Granda2009}. For the curves with $\tilde b>0$ the behaviour changes. In the past $w_D>-1$, but $w_D$ goes down, crosses the phantom divide $w_D=-1$ and enters the phantom region. Later it comes back to $w_D=-1$ from below, which is the de Sitter attractor of Eq.~(\ref{eq:dS}). So for these values KrHDE behaves like quintom DE. A similar crossing followed by a cosmological constant future was found for another generalized entropy in Ref.~\cite{Luciano2026a}. The crossing happens at $z\approx-0.49$, $-0.06$, $0.31$ and $0.52$ for $\tilde b=0.05$, $0.1$, $0.2$ and $0.3$, and Section~\ref{sec:obs} shows that for most of the parameter space allowed by the data it happens only in the far future, or not before $z=-0.9$. The present value $w_{D0}$ moves from $-0.83$ for $\tilde b=0$ to $-0.99$ for $\tilde b=0.1$ and $-1.19$ for $\tilde b=0.3$. In panels (b) and (c), larger $\alpha$ or smaller $\beta$ give a less negative $w_D$ at low redshift.

\begin{figure}[!htbp]
\centering
\includegraphics[width=\textwidth]{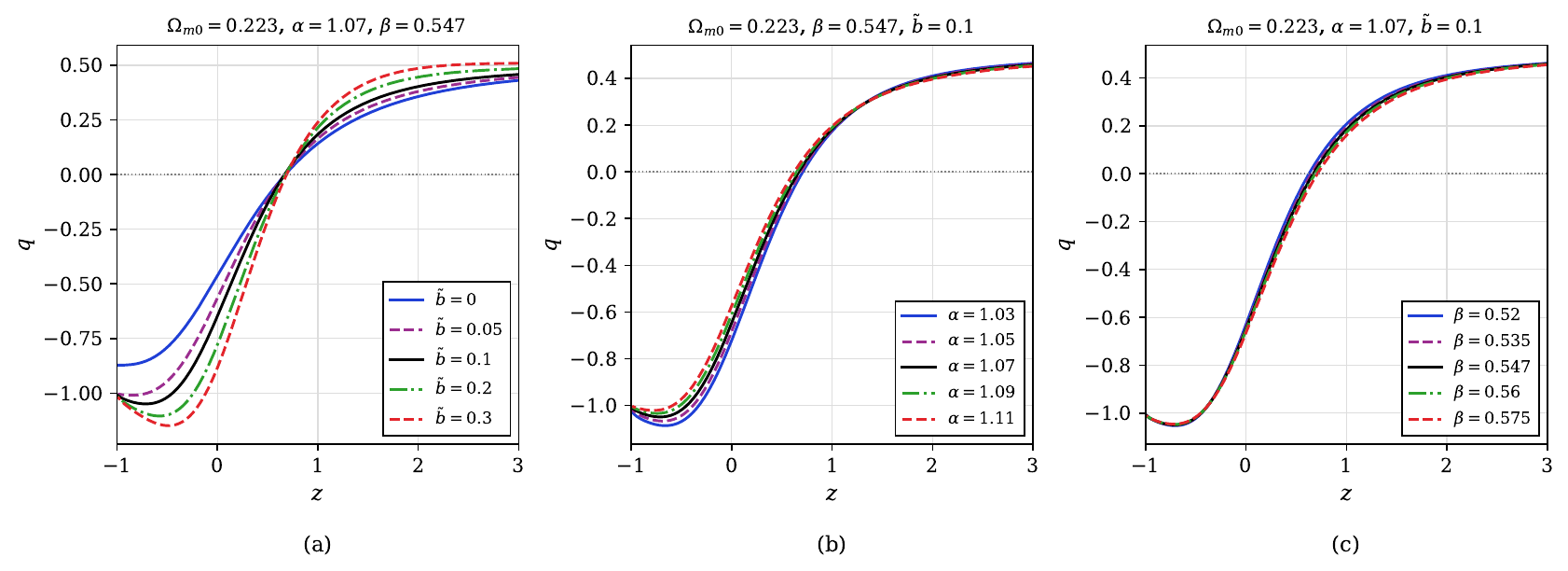}
\caption{The evolution of the deceleration parameter $q$ with redshift $z$. (a) Different $\tilde b$, (b) different $\alpha$, (c) different $\beta$. The dotted line is $q=0$.}
\label{fig:q}
\end{figure}

Fig.~\ref{fig:q} shows the deceleration parameter. In all cases the universe was decelerating in the past ($q\approx0.5$ at $z=3$, the matter era value) and is accelerating now. The transition from deceleration to acceleration happens at $z_t\approx0.67$ for the reference values, and it stays in the range $0.62\lesssim z_t\lesssim0.72$ for all curves. The present value $q_0$ goes from $-0.46$ ($\tilde b=0$) to $-0.65$ ($\tilde b=0.1$) and $-0.89$ ($\tilde b=0.3$). In the far future $q\to-1$ for $\tilde b>0$ (de Sitter), while for $\tilde b=0$ it stays above $-1$.

\begin{figure}[!htbp]
\centering
\includegraphics[width=\textwidth]{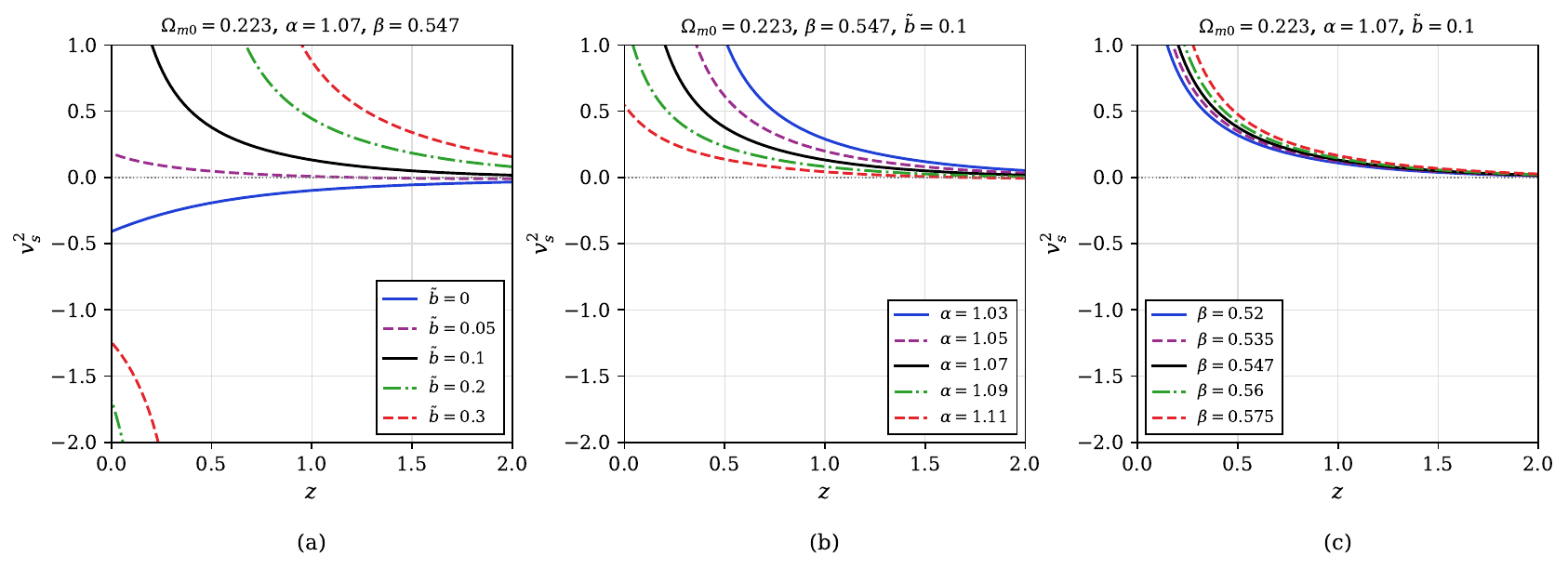}
\caption{The evolution of the squared sound speed $v_s^2$ with redshift $z$. (a) Different $\tilde b$, (b) different $\alpha$, (c) different $\beta$. The dotted line is $v_s^2=0$.}
\label{fig:vs2}
\end{figure}

The squared sound speed is shown in Fig.~\ref{fig:vs2} for $0\le z\le2$. For $\tilde b=0$, $v_s^2<0$ at all redshifts, so the GO-HDE is classically unstable, as found for HDE with other cutoffs \cite{Myung2007,Kim2008}. For $\tilde b\gtrsim0.1$, $v_s^2$ becomes positive at low redshift in the quintessence phase, before the crossing of $w_D=-1$. At higher redshift the Kruglov term becomes small and $v_s^2$ goes back to the negative GO-HDE value (for example, for $\tilde b=0.05$ it is negative for $z>1.2$, and for $\tilde b=0.1$ for $z>2.9$). From Eq.~(\ref{eq:vs2}), $v_s^2$ diverges when $w_D$ crosses $-1$, and after the crossing, in the phantom phase, it is negative (see the curves for $\tilde b=0.2$ and $0.3$ near $z=0$). This is expected, since a single fluid cannot cross $w=-1$ in a stable way \cite{Vikman2005}. So the Kruglov term gives a window of stability at low redshift, but the model is not stable over the whole history. For the parameter values allowed by the data (Section~\ref{sec:obs}), $v_s^2<0$ for almost all of the past evolution, as in GO-HDE.

\section{Observational constraints}
\label{sec:obs}

In this section we find the values of the model parameters from observational data. We use three late-time data sets and one BBN condition.

\subsection{Data sets}

\textbf{Cosmic chronometers (CC).} We use 31 measurements of $H(z)$ in the range $0.07\le z\le1.965$, obtained by the differential age method of passively evolving galaxies \cite{Jimenez2003,Simon2005,Stern2010,Moresco2012,Zhang2014,Moresco2015,Moresco2016,Ratsimbazafy2017}. This method does not depend on any cosmological model. For the point at $z=0.47$ \cite{Ratsimbazafy2017} we add the statistical and systematic errors in quadrature, $\sigma_H=49.6$ km\,s$^{-1}$\,Mpc$^{-1}$. The chi-square is
\begin{equation}
\chi^2_{CC}=\sum_{i=1}^{31}\frac{\left[H_{\rm th}(z_i)-H_{\rm obs}(z_i)\right]^2}{\sigma_{H,i}^2}.
\end{equation}

\textbf{Pantheon+ supernovae (SN).} We use the Pantheon+ sample \cite{Scolnic2022,Brout2022}. We keep the 1590 light curves with $z>0.01$ to avoid strong peculiar velocity effects, and we use the full statistical plus systematic covariance matrix $C_{SN}$. The theoretical distance modulus is $\mu_{\rm th}=5\log_{10}[D_L/{\rm Mpc}]+25$, with
\begin{equation}
D_L=(1+z_{\rm hel})\,\frac{c}{H_0}\int_0^{z_{\rm HD}}\frac{dz'}{E(z')},
\end{equation}
where $z_{\rm HD}$ is the CMB-frame redshift corrected for peculiar velocities and $z_{\rm hel}$ is the heliocentric redshift \cite{Brout2022}.
The absolute magnitude $M$ is not known, so we marginalise over it analytically with a flat prior \cite{Goliath2001,Conley2011}. With $\Delta_i=m_i-\mu_{\rm th}(z_i)$, the chi-square becomes
\begin{equation}
\chi^2_{SN}=\Delta^TC_{SN}^{-1}\Delta-\frac{\left(\sum_{ij}(C_{SN}^{-1})_{ij}\Delta_j\right)^2}{\sum_{ij}(C_{SN}^{-1})_{ij}}.
\end{equation}

\textbf{DESI DR2 BAO.} We use the 13 BAO measurements of DESI DR2 in the range $0.295\le z\le2.33$, with their covariance matrix \cite{DESI2025}. They are given as $D_M/r_d$, $D_H/r_d$ and $D_V/r_d$, where \cite{DESI2025}
\begin{equation}
D_M(z)=\frac{c}{H_0}\int_0^z\frac{dz'}{E(z')},\qquad D_H(z)=\frac{c}{H(z)},\qquad D_V(z)=\left[zD_M^2(z)D_H(z)\right]^{1/3}.
\end{equation}
The sound horizon at the drag epoch, $r_d$, depends on early universe physics. We keep it as a free parameter, so the BAO data constrain only the shape of $H(z)$ and the product $H_0r_d$. The CC data then fix $H_0$. The chi-square is $\chi^2_{BAO}=\Delta^TC_{BAO}^{-1}\Delta$.

\textbf{BBN condition.} As shown in Section~\ref{sec:limits}, the GO cutoff makes the DE follow radiation at early times with the fraction $\Omega_D^{(r)}=\alpha-2\beta$. Using Eq.~(\ref{eq:dNnu}) and the BBN value $N_\nu=2.86\pm0.15$ \cite{Fields2020}, we add
\begin{equation}
\chi^2_{BBN}=\frac{\left(3+\Delta N_\nu-2.86\right)^2}{0.15^2}.
\end{equation}
This term is needed for a second reason. With late-time data alone, $\Omega_{m0}$ and the matter-era fraction $\Omega_D^{(m)}=\alpha-3\beta/2$ cannot be separated. Both scale as $a^{-3}$ at high redshift, so only their combination is measured. For GO-HDE this degeneracy is exact, because the solution of Eq.~(\ref{eq:OmDp}) with $\tilde b=0$ can be found exactly. For $\tilde b=0$ we have $u=E^2-\Omega_{m0}e^{-3x}$, and Eq.~(\ref{eq:E2p}) becomes the linear equation $(E^2)'=(2/\beta)\left[(1-\alpha)E^2-\Omega_{m0}e^{-3x}\right]$. Its solution with $E^2(0)=1$ is the sum of a particular solution $\propto e^{-3x}$ and the homogeneous solution $\propto e^{2(1-\alpha)x/\beta}$, i.e. (see Ref.~\cite{Granda2008}; here written in terms of $\Omega_D^{(m)}$)
\begin{equation}
E^2=\frac{\Omega_{m0}}{1-\Omega_D^{(m)}}(1+z)^3+\left(1-\frac{\Omega_{m0}}{1-\Omega_D^{(m)}}\right)(1+z)^{2(\alpha-1)/\beta}.
\label{eq:GOsol}
\end{equation}
The BBN term fixes $\alpha-2\beta$, and this removes the degeneracy.

The total chi-square is
\begin{equation}
\chi^2_{\rm tot}=\chi^2_{CC}+\chi^2_{SN}+\chi^2_{BAO}+\chi^2_{BBN}.
\end{equation}
For $\Lambda$CDM, the BBN term is zero.

\subsection{Method}

At late times we neglect radiation and solve Eq.~(\ref{eq:E2p}) for $E^2(x)$ from $x=0$ (with $E^2=1$) back to $z=2.4$, using a fourth order Runge--Kutta method. At each step $u$ is found from Eq.~(\ref{eq:root}). This is the same as solving Eq.~(\ref{eq:OmDp}). We explore the parameter space with the affine invariant Markov chain Monte Carlo (MCMC) sampler \texttt{emcee} \cite{Foreman2013}, and we analyse the chains with \texttt{GetDist} \cite{Lewis2019}. The free parameters and their flat priors are given in Table~\ref{tab:priors}. We also require $0\le\Omega_D^{(m)}\le0.3$, so that the DE density stays positive in the matter era and does not dominate there. The same range was used in Ref.~\cite{Sharma2026}. We use 32 walkers, with $8\times10^3$ steps for $\Lambda$CDM, $1.2\times10^4$ steps for GO-HDE and $4\times10^4$ steps for KrHDE, and we remove the first quarter of each chain as burn-in. These chains are longer than 550 times the integrated autocorrelation time. For the KrHDE run with a relaxed prior (Section~\ref{sec:results}) we use $6\times10^4$ steps, which is about 130 autocorrelation times. To compare the models we use the Akaike and Bayesian information criteria \cite{Akaike1974,Schwarz1978,Liddle2007},
\begin{equation}
{\rm AIC}=\chi^2_{\rm min}+2k,\qquad {\rm BIC}=\chi^2_{\rm min}+k\ln N,
\end{equation}
where $k$ is the number of free parameters and $N=1634$ is the number of data points (31 CC, 1590 SN and 13 BAO; counting the BBN term as one more point changes $\Delta$BIC by less than 0.01). A model with smaller AIC or BIC is preferred. A difference smaller than 2 means the models are equally good, while a difference larger than 6 is strong evidence against the model with the larger value \cite{Liddle2007}. We also compute the Bayesian evidence $Z$ with the nested sampling code \texttt{dynesty} \cite{Speagle2020}, using 500 live points and the same priors, and quote $\ln B=\ln Z-\ln Z_{\Lambda{\rm CDM}}$. On the Jeffreys scale, $|\ln B|>5$ is strong evidence \cite{Kass1995}.

\begin{table}[!htbp]
\centering
\caption{Free parameters and their flat priors. $H_0$ is in km\,s$^{-1}$\,Mpc$^{-1}$ and $r_d$ in Mpc.}
\label{tab:priors}
\begin{tabular}{lccc}
\toprule
Parameter & $\Lambda$CDM & GO-HDE & KrHDE \\
\midrule
$H_0$ & $[50,90]$ & $[50,90]$ & $[50,90]$\\
$\Omega_{m0}$ & $[0.05,0.6]$ & $[0.05,0.6]$ & $[0.05,0.6]$\\
$\alpha$ & -- & $[0,3]$ & $[0,3]$\\
$\beta$ & -- & $[0.05,2]$ & $[0.05,2]$\\
$\tilde b$ & -- & 0 (fixed) & $[0,20]$\\
$r_d$ & $[110,190]$ & $[110,190]$ & $[110,190]$\\
\bottomrule
\end{tabular}
\end{table}

\subsection{Results}
\label{sec:results}

\begin{figure}[!htbp]
\centering
\includegraphics[width=0.64\textwidth]{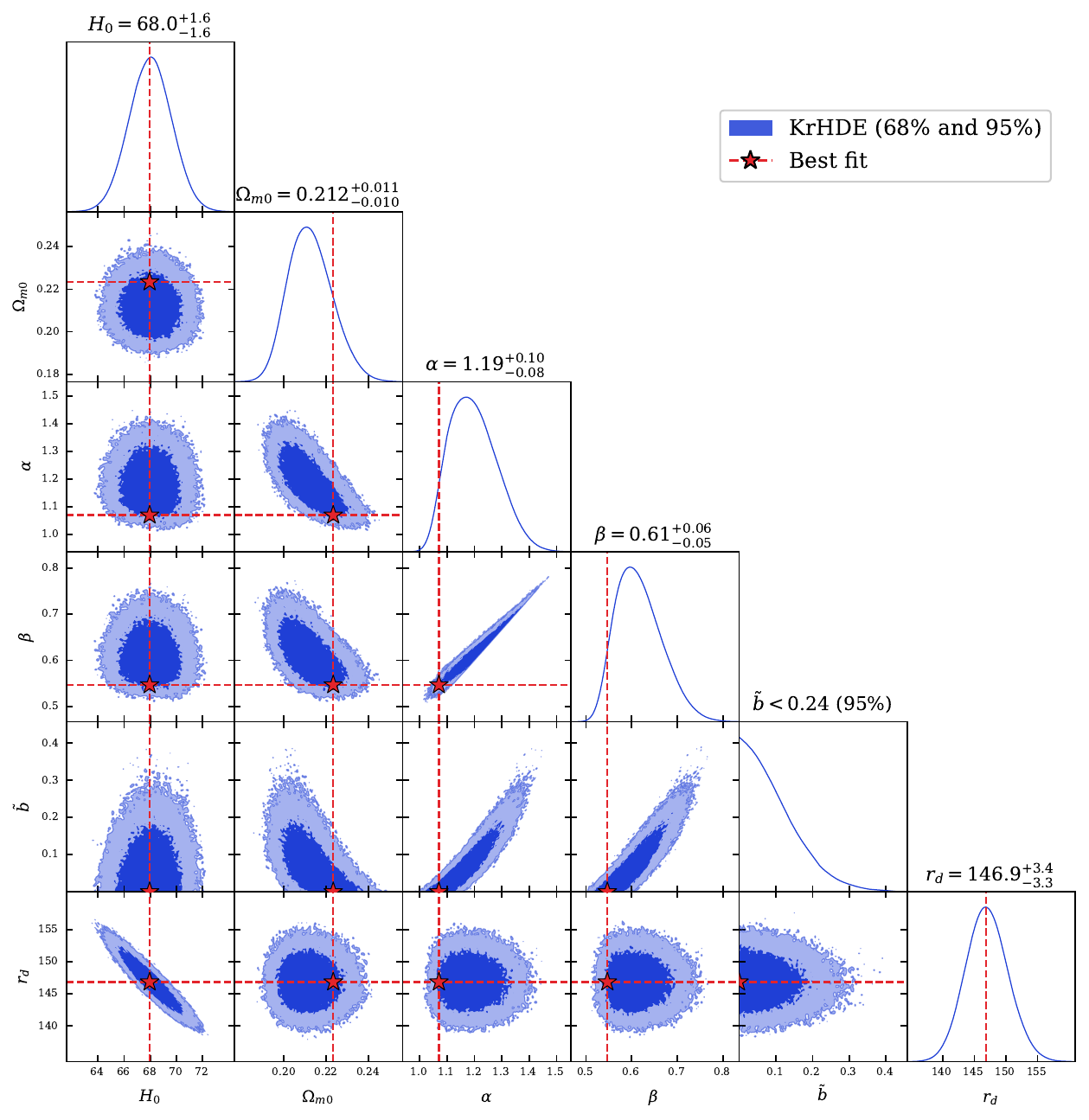}
\caption{One and two dimensional marginalised posteriors of the KrHDE parameters from CC + Pantheon+ + DESI DR2 BAO + BBN. The dark and light regions are the 68\% and 95\% confidence regions. The values above the diagonal panels are the medians with 68\% limits (for $\tilde b$ the 95\% upper limit). The red stars and dashed lines show the best fit point.}
\label{fig:tri}
\end{figure}

The marginalised constraints are given in Table~\ref{tab:post}, and the best fit $\chi^2$ values and information criteria in Table~\ref{tab:ic}. Fig.~\ref{fig:tri} shows the one and two dimensional posteriors of all KrHDE parameters, with 68\% and 95\% confidence regions.

\begin{table}[!htbp]
\centering
\caption{Marginalised constraints (median with 68\% limits). For $\tilde b$ we give the 95\% upper limit. $\Omega_D^{(m)}$ and $\Omega_D^{(r)}$ are derived parameters. $H_0$ is in km\,s$^{-1}$\,Mpc$^{-1}$ and $r_d$ in Mpc.}
\label{tab:post}
\begin{tabular}{lccc}
\toprule
Parameter & $\Lambda$CDM & GO-HDE & KrHDE \\
\midrule
$H_0$ & $68.80^{+1.58}_{-1.61}$ & $67.87^{+1.65}_{-1.63}$ & $68.01^{+1.61}_{-1.62}$\\[2pt]
$\Omega_{m0}$ & $0.3047^{+0.0078}_{-0.0076}$ & $0.2240^{+0.0095}_{-0.0092}$ & $0.2120^{+0.0111}_{-0.0098}$\\[2pt]
$\alpha$ & -- & $1.072^{+0.035}_{-0.034}$ & $1.190^{+0.097}_{-0.082}$\\[2pt]
$\beta$ & -- & $0.549^{+0.023}_{-0.021}$ & $0.612^{+0.057}_{-0.045}$\\[2pt]
$\tilde b$ & -- & 0 & $<0.24$ (95\%)\\[2pt]
$r_d$ & $146.8^{+3.5}_{-3.2}$ & $146.9^{+3.5}_{-3.3}$ & $146.9^{+3.4}_{-3.3}$\\[2pt]
\midrule
$\Omega_D^{(m)}$ & -- & $0.249^{+0.020}_{-0.021}$ & $0.275^{+0.018}_{-0.024}$\\[2pt]
$\Omega_D^{(r)}$ & -- & $-0.026^{+0.024}_{-0.026}$ & $-0.036^{+0.024}_{-0.025}$\\
\bottomrule
\end{tabular}
\end{table}

\begin{table}[!htbp]
\centering
\caption{Best fit $\chi^2$ of each data set, total $\chi^2_{\rm min}$, number of free parameters $k$, AIC and BIC differences, and logarithm of the Bayes factor $\ln B$, all with respect to $\Lambda$CDM ($N=1634$).}
\label{tab:ic}
\small\setlength{\tabcolsep}{4pt}
\begin{tabular}{lccccccccc}
\toprule
Model & $\chi^2_{CC}$ & $\chi^2_{SN}$ & $\chi^2_{BAO}$ & $\chi^2_{BBN}$ & $\chi^2_{\rm min}$ & $k$ & $\Delta$AIC & $\Delta$BIC & $\ln B$\\
\midrule
$\Lambda$CDM & 14.58 & 1405.22 & 10.92 & -- & 1430.72 & 3 & 0 & 0 & 0\\
GO-HDE & 14.71 & 1402.49 & 9.04 & 0.00 & 1426.24 & 5 & $-0.49$ & 10.31 & $-5.9\pm0.3$\\
KrHDE & 14.71 & 1402.49 & 9.04 & 0.00 & 1426.24 & 6 & 1.51 & 17.71 & $-10.8\pm0.3$\\
\bottomrule
\end{tabular}
\end{table}

The main results are as follows.

\begin{figure}[!htbp]
\centering
\includegraphics[width=0.5\textwidth]{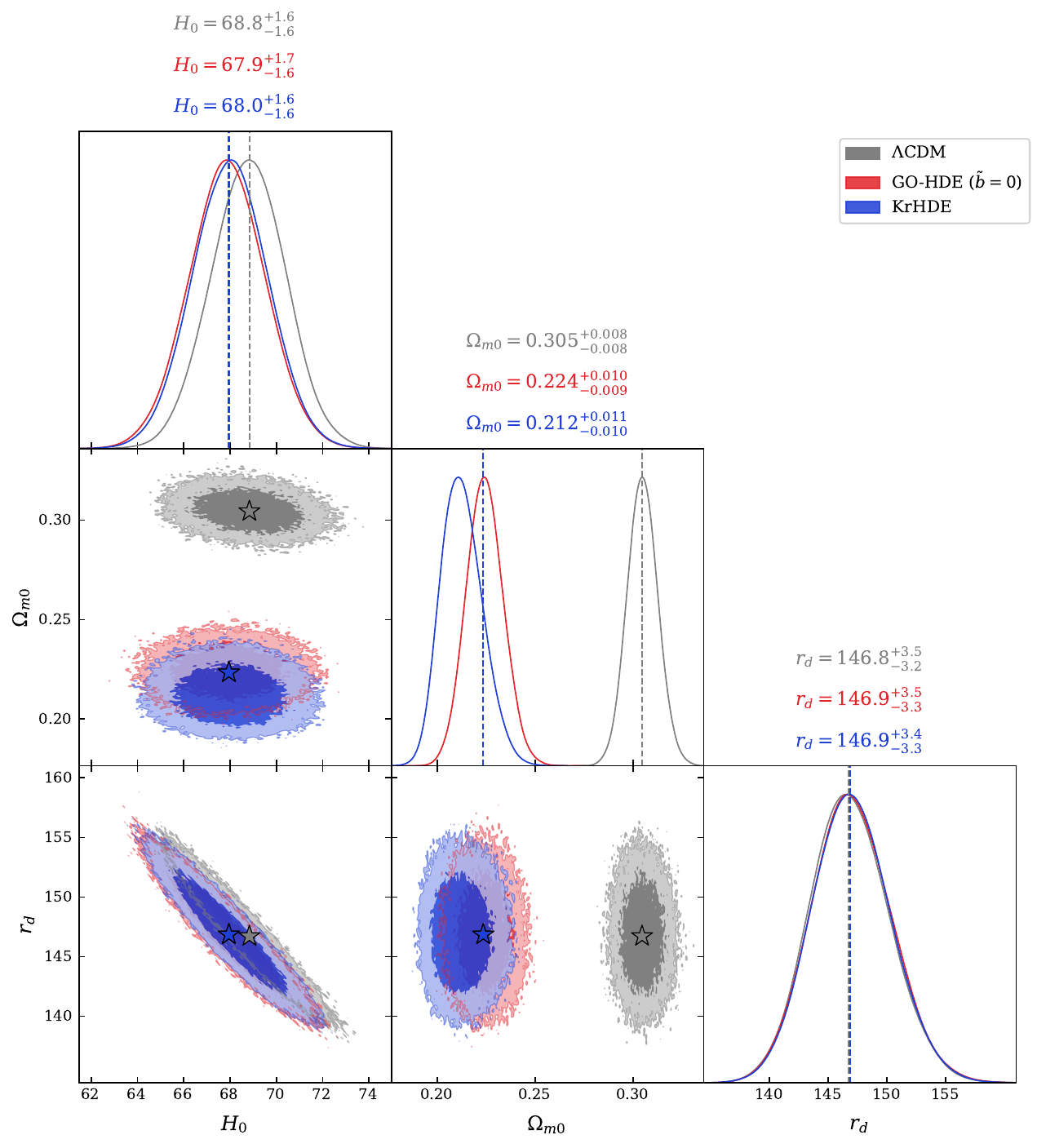}
\caption{Comparison of the constraints on $H_0$, $\Omega_{m0}$ and $r_d$ for $\Lambda$CDM (grey), GO-HDE (red) and KrHDE (blue). The values above the diagonal panels are the medians with 68\% limits for each model. The stars and dashed lines show the best fit of each model.}
\label{fig:comp}
\end{figure}

\begin{figure}[!htbp]
\centering
\includegraphics[width=0.42\textwidth]{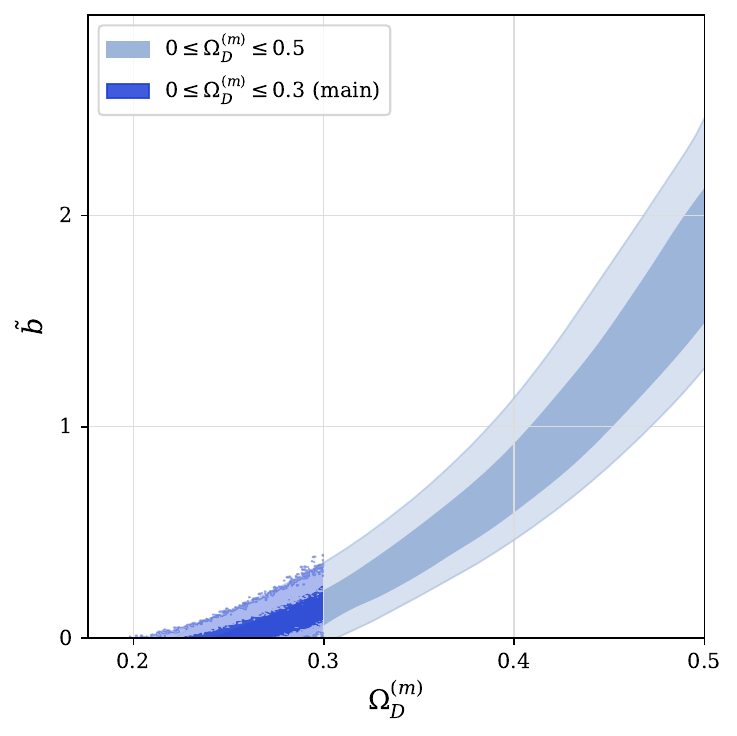}
\caption{Degeneracy between the Kruglov parameter $\tilde b$ and the matter-era DE fraction $\Omega_D^{(m)}$. Dark blue: main analysis with $0\le\Omega_D^{(m)}\le0.3$. Light blue: relaxed prior $0\le\Omega_D^{(m)}\le0.5$.}
\label{fig:degen}
\end{figure}

\begin{enumerate}
\item \textbf{Hubble constant and sound horizon.} All three models give almost the same $H_0\approx68$ km\,s$^{-1}$\,Mpc$^{-1}$ and $r_d\approx147$ Mpc (Fig.~\ref{fig:comp}). Since $r_d$ is free, $H_0$ is fixed mainly by the CC data, and $H_0$ and $r_d$ show the expected strong anti-correlation. The KrHDE does not move $H_0$ towards the local SH0ES value, so it does not solve the Hubble tension.

\item \textbf{Matter density.} The HDE models give a lower $\Omega_{m0}$ ($\approx0.21$--$0.22$) than $\Lambda$CDM ($\approx0.30$). This difference is expected, for the following reasons.
\begin{itemize}
\item With the GO cutoff, a part of the DE behaves like matter at high redshift. From Eq.~(\ref{eq:track}), this fraction is $\Omega_D^{(m)}=\alpha-3\beta/2\approx0.25$ for GO-HDE and $\approx0.28$ for KrHDE (Table~\ref{tab:post}).
\item The late-time data cannot separate this part from the real matter. They measure the total amount that scales as $(1+z)^3$, i.e. the effective matter density [see Eq.~(\ref{eq:GOsol})]
\begin{equation}
\Omega_{m0}^{\rm eff}=\frac{\Omega_{m0}}{1-\Omega_D^{(m)}}.
\label{eq:Omeff}
\end{equation}
\item For GO-HDE, $\Omega_{m0}^{\rm eff}=0.224/(1-0.249)\approx0.30$, and for KrHDE, $\Omega_{m0}^{\rm eff}=0.212/(1-0.275)\approx0.29$. These values agree with the $\Lambda$CDM value $\Omega_{m0}=0.305$.
\end{itemize}
So $\Omega_{m0}$ in the HDE models should not be compared directly with $\Omega_{m0}$ of $\Lambda$CDM; the quantity to compare is $\Omega_{m0}^{\rm eff}$.

\item \textbf{GO parameters.} We find $\alpha>1$ for more than 98\% of the samples in both HDE models. For KrHDE, $\alpha=1.19^{+0.10}_{-0.08}$ and $\beta=0.61^{+0.06}_{-0.05}$. The two parameters are strongly correlated (Fig.~\ref{fig:tri}), and both are larger than in GO-HDE. The BBN term gives $\Omega_D^{(r)}=\alpha-2\beta$ close to zero, so the fit lies near the Ricci line $\alpha=2\beta$. Note that $\Omega_D^{(r)}<0$ for most of the posterior (94\% of the KrHDE samples and 86\% of the GO-HDE samples). This means $L^{-2}<0$ and a small negative DE density in the radiation era. The BBN data allow this only because $|\Omega_D^{(r)}|$ is small, of order a few percent.

\item \textbf{Kruglov parameter.} The best fit is at $\tilde b\simeq0$, and the data give only an upper limit, $\tilde b<0.24$ at 95\% confidence. The data are therefore consistent with the Kruglov correction but do not require it. The posterior of $\tilde b$ is correlated with $\alpha$, $\beta$ and $\Omega_D^{(m)}$, and anti-correlated with $\Omega_{m0}$. A larger $\tilde b$ lowers the late-time DE density (Section~\ref{sec:evol}), and a larger $\alpha$, $\beta$ and early DE fraction can compensate for it. This degeneracy is shown in Fig.~\ref{fig:degen}. If we relax the prior to $0\le\Omega_D^{(m)}\le0.5$, the allowed region extends along a narrow band to large $\tilde b$, and we get $\tilde b=0.96^{+0.63}_{-0.66}$ and $\tilde b<1.90$ (95\%). Points with $\tilde b\approx1.7$ and $\Omega_D^{(m)}\approx0.5$ have $\chi^2$ only about 1 larger than the best fit. So the bound on $\tilde b$ depends on how much early DE is allowed. An independent limit on early DE, for example from the CMB, would make the bound on $\tilde b$ much stronger.

\item \textbf{Entropy parameter.} The limit $\tilde b<0.24$ means $b<0.24H_0^4\approx1.1\times10^{-168}$ GeV$^4$. Using $b=\pi^2\gamma/G^2=64\pi^4\gamma M_p^4$, this gives $\gamma<4.9\times10^{-246}$. For comparison, Ref.~\cite{Phukan2026} found $b<9.4\times10^{-105}$ GeV$^4$ from BBN for the same entropy, but in the apparent horizon cosmology of Ref.~\cite{Kruglov2025} (they call the parameter $\lambda$). Our late-time limit is much smaller, but the two limits refer to different models and cannot be compared directly. In all accepted samples $u_0^2>\tilde b$, so today the horizon entropy is on the increasing part of $S_K$, i.e. $\gamma S_{BH}^2<1$. Also $1<\alpha<2$ for 99.9\% of the KrHDE samples, so the future de Sitter state of Eq.~(\ref{eq:dS}) is on the same branch. The final Hubble rate is $H_{dS}=0.74^{+0.06}_{-0.11}\,H_0$.
\end{enumerate}

Comparing the two HDE models in Table~\ref{tab:post}, adding $\tilde b$ widens the ranges of $\alpha$ and $\beta$ and moves them to larger values, and the posterior of $\Omega_D^{(m)}$ moves towards the edge of the prior (its 95\% upper value is 0.298). This is the same degeneracy as in Fig.~\ref{fig:degen}.

\textbf{Model comparison.} Both HDE models fit the data better than $\Lambda$CDM, with $\Delta\chi^2=-4.49$. Most of the gain comes from the supernovae ($-2.73$) and the BAO data ($-1.88$). The KrHDE gives the same $\chi^2_{\rm min}$ as GO-HDE, because its best fit is at $\tilde b\simeq0$. After the penalty for extra parameters, GO-HDE has $\Delta{\rm AIC}=-0.49$ and KrHDE has $\Delta{\rm AIC}=1.51$. Both differences are smaller than 2, so by AIC all three models are equally supported. The BIC, which penalises extra parameters more strongly for a large data set, gives $\Delta{\rm BIC}=10.31$ for GO-HDE and $17.71$ for KrHDE, which is strong evidence in favour of $\Lambda$CDM. The Bayesian evidence gives the same result: $\ln B=-5.9\pm0.3$ for GO-HDE and $-10.8\pm0.3$ for KrHDE, which is strong evidence for $\Lambda$CDM on the Jeffreys scale \cite{Kass1995}. The evidence depends on the prior ranges of $\alpha$, $\beta$ and $\tilde b$; most of the penalty comes from the large prior volume of these parameters, compared with the small region allowed by the data. So the present data do not need the extra freedom of the HDE models, and in particular they do not need the Kruglov parameter.

\begin{figure}[!htbp]
\centering
\includegraphics[width=\textwidth]{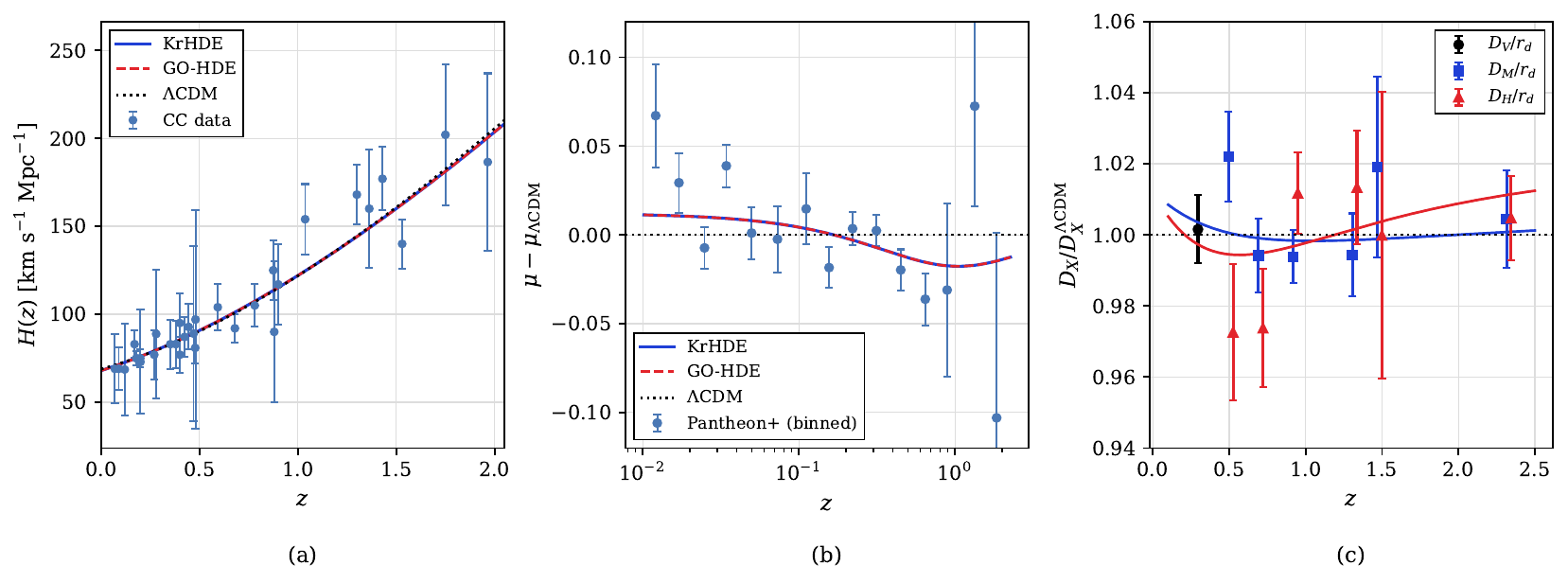}
\caption{Best fit models compared with the data. (a) $H(z)$ with the 31 CC points. (b) Distance modulus relative to the best fit $\Lambda$CDM, with Pantheon+ data in redshift bins (each model uses its own best fit $M$). (c) DESI DR2 BAO distances divided by the best fit $\Lambda$CDM prediction. Solid lines: KrHDE; dashed lines: GO-HDE. The KrHDE best fit has $\tilde b\simeq0$, so its curves lie on top of the GO-HDE curves.}
\label{fig:data}
\end{figure}

Fig.~\ref{fig:data} compares the best fit models with the data. All models pass through the CC points (panel a). In panel (b), the HDE models are slightly above $\Lambda$CDM at low redshift and below it near $z\approx1$, which follows the trend of the binned supernova data. In panel (c), the HDE models predict a smaller $D_H/r_d$ near $z\approx0.5$--$0.7$, where the DESI data also lie below $\Lambda$CDM. These small differences explain the lower $\chi^2$ of the HDE models.

\begin{figure}[!htbp]
\centering
\includegraphics[width=\textwidth]{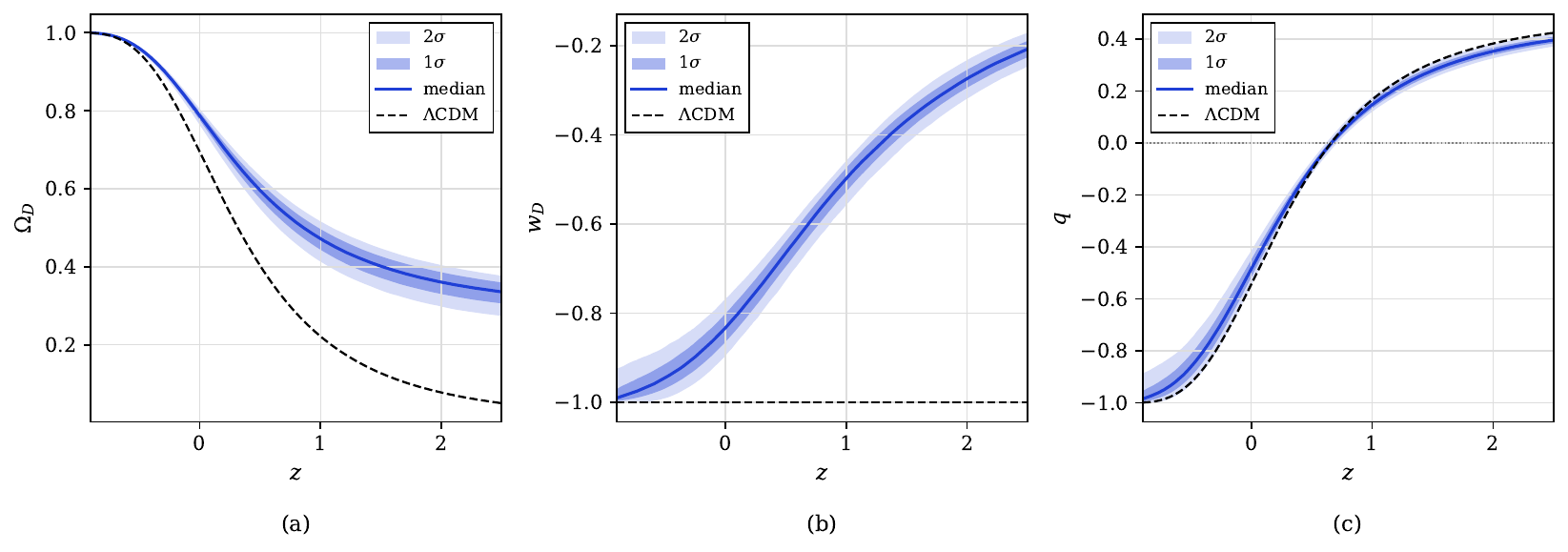}
\caption{Reconstructed (a) $\Omega_D(z)$, (b) $w_D(z)$ and (c) $q(z)$ of KrHDE from the MCMC chain, with 68\% and 95\% bands. The dashed lines show the best fit $\Lambda$CDM.}
\label{fig:rec}
\end{figure}

Finally, Fig.~\ref{fig:rec} shows $\Omega_D$, $w_D$ and $q$ reconstructed from the chain. We find $w_{D0}=-0.833\pm0.031$, so today the KrHDE behaves like quintessence. This agrees with the DESI DR2 preference for $w_0>-1$ \cite{DESI2025}. However, in our model $w_D>-1$ at all past redshifts, while the DESI $w_0w_a$ fits prefer $w<-1$ at higher redshift. The deceleration parameter today is $q_0=-0.484^{+0.036}_{-0.032}$, and the transition redshift is $z_t=0.659^{+0.027}_{-0.021}$, practically the same as in $\Lambda$CDM ($z_t\approx0.66$). In the future $w_D\to-1$ and $q\to-1$, as expected from the de Sitter attractor. For most of the allowed region the phantom crossing does not happen before $z=-0.9$; only about 5\% of the samples cross $w_D=-1$, all in the future ($-0.88<z<-0.39$). At $z=2.5$ we find $\Omega_D=0.34^{+0.02}_{-0.03}$, and it decreases slowly towards $\Omega_D^{(m)}\approx0.28$ at higher redshift.

We also checked the classical stability over the posterior. The squared sound speed today is positive for only 5\% of the samples, and for 99.9\% of the samples $v_s^2<0$ somewhere at $z>2$. So, for the parameters allowed by the data, KrHDE is classically unstable for almost all of its past evolution, like GO-HDE.

\section{Summary of results}
\label{sec:summary}

In this paper we built a holographic dark energy model from the Kruglov entropy $S_K=S_{BH}/(1+\gamma S_{BH}^2)$ with the Granda--Oliveros cutoff $L^{-2}=\alpha H^2+\beta\dot H$ in a flat universe. We called it KrHDE. The Kruglov entropy is smaller than the BH entropy and has a maximum value, which makes it different from the Tsallis, Barrow, R\'enyi and Kaniadakis entropies. The main results are given below.

\begin{enumerate}
\item The energy density is $\rho_D=3M_p^2X^3/(X^2+b)$ with $X=\alpha H^2+\beta\dot H$ and $b=\pi^2\gamma/G^2$. The Friedmann equation gives a cubic equation for the cutoff, which has one real root. Using this root, we obtained the differential equation (\ref{eq:OmDp}) for $\Omega_D$, and the expressions for $w_D$, $q$ and $v_s^2$. For $\tilde b=0$ all results reduce to those of GO-HDE.

\item At early times the Kruglov correction is negligible, and the DE follows the background with the fractions $\alpha-2\beta$ in the radiation era and $\alpha-3\beta/2$ in the matter era. At late times the correction becomes important. For $\alpha>1$ the model has a stable de Sitter future with $H_{dS}^4=b/[\alpha^2(\alpha-1)]$, so the final Hubble rate is fixed by the entropy parameter. GO-HDE has no such state.

\item The model gives a matter era followed by a DE era, with a transition from deceleration to acceleration at $z_t\approx0.66$. For larger $\tilde b$ the EoS parameter crosses the phantom divide and then approaches $-1$, so the model behaves like quintom DE. The squared sound speed becomes positive at low redshift for $\tilde b\gtrsim0.1$, but it is negative at higher redshift and in the phantom phase. For the parameters allowed by the data the model is classically unstable for almost all of its past, like GO-HDE.

\item Using CC, Pantheon+ and DESI DR2 BAO data together with a BBN condition, we found $H_0=68.0\pm1.6$ km\,s$^{-1}$\,Mpc$^{-1}$, $\Omega_{m0}=0.212^{+0.011}_{-0.010}$, $\alpha=1.19^{+0.10}_{-0.08}$, $\beta=0.61^{+0.06}_{-0.05}$ and $\tilde b<0.24$ at 95\% confidence, i.e. $\gamma<4.9\times10^{-246}$. The best fit is at $\tilde b\simeq0$. Today $w_{D0}=-0.83\pm0.03$ and $q_0=-0.48\pm0.03$. Phantom crossing, if it happens, is in the future.

\item The bound on $\tilde b$ is degenerate with the early DE fraction $\Omega_D^{(m)}$. If up to 50\% early DE is allowed in the matter era, values of $\tilde b$ up to about 1.9 are also allowed.

\item KrHDE and GO-HDE fit the data slightly better than $\Lambda$CDM ($\Delta\chi^2=-4.5$), and by AIC the three models are equally good. But BIC and the Bayesian evidence prefer $\Lambda$CDM strongly, because the HDE models have more parameters.
\end{enumerate}

The Kruglov HDE with GO cutoff is consistent with the present late-time data and has two new features: a de Sitter future set by the entropy parameter and a possible phantom crossing. However, the present data do not require the Kruglov correction. Two steps can improve the analysis. The first is to add CMB information, which limits the early DE fraction and so breaks the degeneracy between $\tilde b$ and $\Omega_D^{(m)}$. The second is to study the growth of perturbations, which is sensitive to the early DE. The extension of this model to a non-flat universe is also of interest, and we leave these studies for future work.

\section*{CRediT authorship contribution statement}
\textbf{Umesh Kumar Sharma:} Conceptualization, Supervision, Writing -- review \& editing. \textbf{P. Suresh Kumar:} Methodology, Software, Formal analysis, Visualization, Writing -- original draft.

\section*{Declaration of competing interest}
The authors declare that they have no known competing financial interests or personal relationships that could have appeared to influence the work reported in this paper.

\section*{Funding}
This research did not receive any specific grant from funding agencies in the public, commercial, or not-for-profit sectors.

\section*{Data availability}
The data used in this work are public: the cosmic chronometer compilation listed in Refs.~\cite{Jimenez2003,Simon2005,Stern2010,Moresco2012,Zhang2014,Moresco2015,Moresco2016,Ratsimbazafy2017}, the Pantheon+ release \cite{Scolnic2022,Brout2022} and the DESI DR2 BAO release \cite{DESI2025}.

\appendix
\section{Root of the cubic equation}
\label{app:cardano}

Here we give the steps that lead to Eq.~(\ref{eq:root}). We start from the cubic (\ref{eq:cubic}),
\begin{equation}
u^3-\mathcal{D}u^2-\mathcal{D}\tilde b=0 .
\end{equation}
\textbf{Step 1.} We remove the $u^2$ term by the substitution $u=t+\mathcal{D}/3$. Expanding $(t+\mathcal{D}/3)^3-\mathcal{D}(t+\mathcal{D}/3)^2-\mathcal{D}\tilde b$, the $t^2$ terms cancel and we get the reduced cubic
\begin{equation}
t^3+pt+q=0,\qquad p=-\frac{\mathcal{D}^2}{3},\qquad q=-\left(\frac{2\mathcal{D}^3}{27}+\mathcal{D}\tilde b\right).
\end{equation}
\textbf{Step 2.} The discriminant of the reduced cubic is
\begin{equation}
\Delta=\frac{q^2}{4}+\frac{p^3}{27}=\left(\frac{\mathcal{D}^3}{27}+\frac{\mathcal{D}\tilde b}{2}\right)^2-\frac{\mathcal{D}^6}{729}=\frac{\mathcal{D}^4\tilde b}{27}+\frac{\mathcal{D}^2\tilde b^2}{4}\ \geq 0 .
\end{equation}
Since $\Delta\geq0$, there is exactly one real root, in agreement with the fact that $u^3/(u^2+\tilde b)$ is an increasing function.

\textbf{Step 3.} By Cardano's formula the real root is $t=\mathcal{A}+\mathcal{B}$, with
\begin{equation}
\mathcal{A}=\left[-\frac{q}{2}+\sqrt{\Delta}\right]^{1/3},\qquad \mathcal{B}=\left[-\frac{q}{2}-\sqrt{\Delta}\right]^{1/3}.
\end{equation}
The product is $\mathcal{A}\mathcal{B}=(q^2/4-\Delta)^{1/3}=(-p^3/27)^{1/3}=\mathcal{D}^2/9$, so $\mathcal{B}=\mathcal{D}^2/(9\mathcal{A})$.

\textbf{Step 4.} Going back to $u=t+\mathcal{D}/3$, we obtain
\begin{equation}
u=\frac{\mathcal{D}}{3}+\mathcal{A}+\frac{\mathcal{D}^2}{9\mathcal{A}},\qquad
\mathcal{A}=\left[\frac{\mathcal{D}^3}{27}+\frac{\mathcal{D}\tilde b}{2}+\sqrt{\frac{\mathcal{D}^4\tilde b}{27}+\frac{\mathcal{D}^2\tilde b^2}{4}}\,\right]^{1/3},
\end{equation}
which is Eq.~(\ref{eq:root}). For $\tilde b=0$, $\Delta=0$ and $\mathcal{A}=\mathcal{D}/3$, so $u=\mathcal{D}$. In the numerical work, when $\mathcal{D}<0$ we first compute $\mathcal{B}$ and then $\mathcal{A}=\mathcal{D}^2/(9\mathcal{B})$, which avoids loss of accuracy.

\end{document}